# Generalizable automated ischaemic stroke lesion segmentation with vision transformers

Chris Foulon*[1], Robert Gray[1], James K Ruffle[1,2], Jonathan Best[1], Tianbo Xu[1], Henry Watkins[1], Jane Rondina[1], Guilherme Pombo[1], Dominic Giles[1], Paul Wright[3], Marcela Ovando-Tellez[4], H. Rolf Jäger[2], Jorge Cardoso[3], Sebastien Ourselin[3], Geraint Rees[1] and Parashkev Nachev*[1].

[1]*UCL Queen Square Institute of Neurology, University College London, London, UK*
[2]*Lysholm Department of Neuroradiology, National Hospital for Neurology and Neurosurgery, London, UK*
[3]*School of Biomedical Engineering and Imaging Sciences, King's College London, London, UK*
[4]*Groupe d'Imagerie Neurofonctionnelle, Institut des Maladies Neurodégénératives-UMR 5293, CNRS, CEA, University of Bordeaux, Bordeaux, France*
*Corresponding authors*

## Abstract

Ischaemic stroke, a leading cause of death and disability, critically relies on neuroimaging for characterising the anatomical pattern of injury. Diffusion-weighted imaging (DWI) provides the highest expressivity in ischemic stroke but poses substantial challenges for automated lesion segmentation: susceptibility artefacts, morphological heterogeneity, age-related comorbidities, time-dependent signal dynamics, instrumental variability, and limited labelled data. Current U-Net-based models therefore underperform, a problem accentuated by inadequate evaluation metrics that focus on mean performance, neglecting anatomical, subpopulation, and acquisition-dependent variability. Here, we present a high-performance DWI lesion segmentation tool addressing these challenges through optimized vision transformer-based architectures, integration of 3563 annotated lesions from multi-site data, and algorithmic enhancements, achieving state-of-the-art results. We further propose a novel evaluative framework assessing model fidelity, equity (across demographics and lesion subtypes), anatomical precision, and robustness to instrumental variability, promoting clinical and research utility. This work advances stroke imaging by reconciling model expressivity with domain-specific challenges and redefining performance benchmarks to prioritize equity and generalizability, critical for personalized medicine and mechanistic research.

# Introduction

Stroke is globally the leading cause of adult neurological disability and the second commonest cause of death[1]. Neuroimaging is central to characterising the underlying pathological process and revealing the lesioned neuroanatomy on which subsequent functional deficits critically depend. Delineation of characteristic phenotypes[2], prediction of clinical outcomes[3], prescription of optimal treatments[4], and inference to modifiable disease mechanisms[5] all rely on imaging-derived representations of the stroke-injured brain and are potentially enhanced by robust, accurate, high-resolution, objective, reproducible means of representation only an automated process could conceivably deliver. Indeed, until the resolving power of an anatomical representation of stroke matches the resolution of the brain's functional anatomy, including its inter-patient variation—a practical impossibility—*any* model of stroke is bound to fall short of the optimum: the ideal can be only asymptotically approached. Given that the avowed objective of medicine is to realise everyone's potential for health at the individual level, regardless of background, our primary task is not to deliver a definitive image-analytic tool but to define a modelling framework that prescribes an optimal trajectory of analytic model improvement, fully cognizant of the distinctive challenges of the domain and the critical aspects of achieved performance.

In ischaemic stroke, the commonest kind, one imaging modality—diffusion-weighted imaging (DWI)—currently offers the best anatomically specific signal in the all-important acute phase of injury[6]. Though lesion contrast is high, both against healthy tissue and rival pathologies, six characteristics of DWI and the applicable clinical context hinder the task of extracting lesion representations with high fidelity. First, images are commonly corrupted by magnetic susceptibility artefacts within the intensity distribution of lesions that exhibit a complex spatial structure arising from the interaction of incidental anatomical and instrumental features[7]. Whether a focal region of signal abnormality is a true ischaemic lesion is strongly modulated not only by anatomical location but also by structured inter-individual variation that needs a highly expressive model to capture. Second, ischaemic lesions exhibit marked morphological heterogeneity[8], reflecting a complex generative process dependent on vascular topology, mechanisms of occlusion, and the clinical eloquence of the lesion. Differences in the frequency of morphological subtypes promote variability in their characterisation, resulting in morphology-specific differences in fidelity. Third, the parenchymal background on which lesion signals are superimposed is both modulated by common age-related comorbidities such as chronic cerebrovascular disease and suboptimally conveyed by the weak normal tissue contrast of DWI. The acute lesion signal is, therefore, hard to contextualise appropriately. Fourth, the relation between the underlying pathological process and the DWI is time-dependent, complicating the interpretation of the signal where the onset of stroke is uncertain and rendering desirable the incorporation of clinical features not routinely stored with imaging data. Fifth, since DWI protocols vary widely in their acquisition parameters and involve comparatively complex physics, a wide range of instrumental variability must be learnt directly from the data, disentangled from other sources of variation, without the aid of physics-derived priors. Finally, the high model expressivity the foregoing characteristics imply must be delivered within the modest scale data regimes that reign in the domain of clinical neuroimaging, with especially tight constraints on the availability of densely labelled images.

These characteristics interact to make arguably the most basic image analytic task here—dense lesion segmentation—far more challenging than casual inspection of DWI images in stroke may suggest. The difficulty perhaps explains the relative paucity of DWI-based lesion segmentation models[9–13] and their modest performance compared with kindred tasks such as brain tumour segmentation[14–17] informed by smaller scale data. It also invites reflection on whether refinement of the established approach to lesion segmentation, where U-Net-based models dominate[18,19], is sufficient or whether a radical change in strategy is required.

But the challenge extends beyond model architectural development to the criteria used to evaluate performance. The dominant evaluative approach—quantification of mean performance on a sample test set—is fundamentally at odds with the demands of the downstream tasks representations of neuroimaging are intended to serve. It assumes, first, that mean image-level performance is a sufficient descriptor of fidelity, ignoring heterogeneities across the population; second, that anatomical, voxel-level differences in fidelity are immaterial, treating lesions as interchangeable; and third, that other contextual systematic variation across samples, including instrumental effects, can be ignored. Clearly, subpopulation-level differences of any source—biological, pathological, or instrumental—introduce inequity in any downstream model, the clinical and mechanistic consequences of lesions depend on their voxel-level features, and generalisability across acquisition settings is crucial to both clinical and research uses. Development—and ultimately real-world application—require far richer indices of model performance if the benefits of image analysis are to be realised in the clinical and scientific realms.

In the present study, we design and implement an approach to creating a high-performance acute ischaemic stroke segmentation tool for DWI that directly addresses the multiple challenges of the task, drawing on current model architectures optimally suited to the problem, large-scale, multi-site imaging data, and high-performance computing. We introduce an array of algorithmic and training enhancements that improve the fidelity, equity, robustness, and generalisability of the resultant model and achieve state-of-the-art performance on conventional metrics. Furthermore, we propose a new, comprehensive evaluative framework that enables the assessment of any candidate model with respect to subpopulation equity, lesion anatomy, and instrumental quality, where each may be richly defined so that utility in clinical and research settings can be adequately corroborated[20]. We make all code available open source.

# Material and Methods

## 2.1 Data

### 2.1.1 Data description
Our objective is to develop a lesion segmentation model for DWI in confirmed acute ischaemic stroke, operable across diverse populations, patterns of ischaemic injury, and scanner parameters. This requires maximising the volume, inclusivity, and instrumental variability of training and testing data to the limit of feasibility. We employed several complementary datasets obtained from routine clinical care—along with a subset from the UK Biobank—that varied in their originating institution and timing but shared identical enrolment criteria. The inclusion criteria were a documented diagnosis of acute ischaemic stroke with DWI performed within 14 days of admission (for positive cases) and a radiologically reported absence of ischaemic injury on DWI (for negative cases). The exclusion criteria, established from the accompanying radiological report and manual curation, were the presence of other pathology manifesting as focal high signal on DWI or gross artefact, such as from large patient movements, other than standard susceptibility artefact, that rendered the image unreportable. Dataset 1 consisted of 1333 previously reported DWI-positive imaging studies from University College London Hospitals NHS Trust (UCLH) of patients with ischaemic stroke[8,21], for which semi-automated manually curated lesion masks were available[22]. Dataset 2 consisted of 5139 previously unreported DWI-positive studies from UCLH with a clinical diagnosis of acute ischaemic stroke. Dataset 3 consisted of 6691 DWI-negative studies from UCLH patients investigated for a wide range of indications. Dataset 4 consisted of 1000 randomly selected participants in UK Biobank for whom DWI was available [23,24]. Dataset 5 consisted of 2674 previously unreported DWI-positive studies from King's College University Hospitals NHS Trust (KCH) with a clinical diagnosis of acute ischaemic stroke. To minimise the impact of clinical and administrative errors in labelling, each image was directly inspected by experts trained in neurology (PN or JB) or neuroradiology (JR).

After curation (summarised in Figure 1), we combined the curated images from Datasets 1 and 2 into a single "Training Lesions" dataset (n = 3563) with corresponding lesion labels (Figure 2A) and the curated controls from Datasets 3 and 4 into a "Training Controls" dataset (n = 6900). We used these datasets for 5-fold cross-validation, whereby each fold used 80% of the images for training and 20% for validation/testing. To ensure a representative distribution of lesions across the five folds, we balanced each split statistically by lesion location and lesion size before training so that particular lesion patterns dominated no single fold. This approach (described in Section 2.3.2) ensures each image is evaluated in a held-out manner exactly once, providing a robust estimate of out-of-sample performance. Because Dataset 5 (n = 2674) has no expert lesion labels, we did not include it in formal training or validation metrics; instead, we used it exclusively to verify that the distribution of predicted lesions (Figure 3) was comparable to that seen in our internal data.

### 2.1.2 Demographics

The final training lesions set of positive studies (n = 3563) included patients with a mean age of 66.94 (standard deviation [SD] 15.24) years old (42.67% of females), and the final training controls set of negative studies (n = 6900) included patients with a mean age of 45.31 (SD 15.16) years old (62.62% of females). The KCH external lesions set of positive studies included patients with a mean age of 65.96 (SD 15.42) years old (41.51% of females). The five training folds had a mean patient age of 66.94 ± 0.50 years (within-fold SD: 15.24 ± 0.19) and a mean proportion of female patients of 43% ± 3% (within-fold SD: 0.49 ± 0.01). Lesion sizes averaged 1327.58 ± 6.94 (within-fold SD: 3065.98 ± 57.60). More detailed descriptive statistics are available in Supplementary Table 1.

### 2.1.3 Imaging parameters

All imaging, except for Dataset 4, was conducted in the course of routine clinical care over a period of 18 years on several different scanners and with a wide variety of acquisition protocols. The scanners included GE Medical Systems, Philips Healthcare & Medical Systems, and Siemens. The clinical DWI sequences consisted of 2 to 16 b-values / b-vectors (only b-values of 0 and 1000 were used during preprocessing) per acquisition session. The image resolution ranged from 128 to 384 voxels in the x-axis, 30 to 384 in y and 19 to 256 in z. The Dataset 4 from the UK Biobank[25] provides standardised DWI images with a resolution of 2 x 2 x 2 mm (104 x 104 x 72 voxels) and five images with a b-value of 0 and 50 with a b-value of 1000 acquired on Siemens 3T machines.

### 2.1.4 DWI preprocessing

The same pre-processing steps implemented in SPM12[26] were applied to all images. For each imaging study, we first computed the geometric mean of the images with a b-value of 0 (b0) and the geometric mean of those with a b-value of 1000 (b1000). Second, the b0 and b1000 mean images were rigid-aligned to MNI space using SPM's rigid registration algorithm. Third, the parameters for a non-linear registration to MNI were obtained from the rigid-aligned mean b0 and applied to it and to the rigid-aligned mean b1000, resliced to 2x2x2mm isotropic resolution, non-linearly transforming both images in MNI space. SPM12's default parameters for non-linear normalization were used. Non-linear registration was used to achieve conformity with the anatomical space of the labels in Dataset 1 and to facilitate analysis of the spatial distribution of lesions. We then applied additional transformations to accommodate the input requirements of our deep-learning models. Every image and label, where available, was transformed to fit the training model's dimensions. We first pad the images, labels and controls from the standard 91x109x91 SPM space to the 96x128x96 resolution of the architecture. We chose padding instead of the common resizing to avoid unnecessary interpolation to the original data. We then normalise the intensity of the images and controls by removing the mean and rescaling it between 0 and 1 to transform the data in a bounded interval. This normalisation is performed before and after data augmentations to ensure the model sees comparable data. Finally, we use the CoordConv method[27], in which we add three channels to the DWI image channel, each containing the coordinates in one axis—0 to 95 in the x-direction, 0 to 127 in the y-direction and 0 to 95 in the z-direction. CoordConv has been shown to help models handle spatial transformations in the context of anatomically sensitive signals.

### 2.1.5 Label curation

Manual dense segmentation labelling is infeasible at this data scale. We therefore adopted an iterative approach, where machine-generated segmentations were manually selected and, where appropriate, modified to conform to the expert-perceived ground truth and used to train successive models. The initial machine-generated segmentations available for Dataset 1 were those previously derived with the zeta anomaly-based unsupervised method described in Mah et al., 2014, following manual curation, and reported elsewhere[8,21]. The first iteration employed the MONAI[28] implementation of a residual U-Net[29] and was used to resegment the initial training dataset (Dataset 1) and the part of Dataset 2 that was available at the time. The subsequent curation stage yielded 2429 labelled studies judged to be of sufficient quality to be used as ground truth for further training. The second iteration employed UNETR[30], selected for its theoretically superior ability to learn long-range associations between signals of the kind needed to distinguish lesions from artefacts, and was again applied to the whole of Datasets 1 & 2. The manual curation stage yielded a total of 3563 satisfactorily labelled images. The voxel-wise spatial distribution of these ground truth dense lesion labels is displayed in Figure 2A.

## 2.2 Model design

### 2.2.1 Architecture

For our final model, we chose the SWIN-UNETR architecture[31], an updated version of UNETR with better performance on benchmark tests. We trained two SWIN-UNETR models: one with DWI-negative controls added to the training data (SWIN-UNETR+Ctr) and one without (SWIN-UNETR). We also trained a simple U-Net without data augmentation as a low-level baseline. This allowed us to explore the differences between model architectures and the more subtle differences in adding controls to the training set independently.

### 2.2.2 Data augmentation

In line with established practice, data augmentation was used to minimise overfitting and improve generalisability[32]. We used three categories of data augmentations randomly applied to the input using the MONAI Python library[28]. The probability of applying the augmentation on a given image, chosen empirically between 5, 10 and 20%, depended on the nature and computational intensity of the augmentation. We used signal intensity operations—intensity shifts and histogram shifts—to prevent the model from focusing too much on a specific range and potentially ignoring lesions with uncommon patterns. We also used geometric operations to account for natural spatial variations—we randomly applied rotations, shears, translations, scaling and hemisphere flip (flip of the x-axis). Finally, we incorporated different varieties of noise that can corrupt DWI acquisitions: Gibbs noise[33], Rician noise[34], Spike noise[35], and bias fields[36].

### 2.2.3 Loss functions

Training a model requires carefully choosing the loss function(s) that will penalise its errors in a task-relevant way. We used the Dice loss[37], which is widely used in segmentation models. The Dice score quantifies the intersection between the predicted segmentation and the ground truth. Unfortunately, this loss is insensitive when lesion size is small. To alleviate this drawback, we sum this Dice loss with the Focal loss[38], a variation of the binary Cross-Entropy classification loss—here classifying voxels as lesioned or not. The Focal loss has been designed to handle large class imbalance, such as the vast majority of voxels not being lesioned. We used a focusing parameter of 2 for the Focal loss (gamma in the MONAI implementation and λ in the Lin et al. study[38]).

### 2.2.4 Control false positive loss

Standard loss functions such as the Dice and the Hausdorff Distance[39] (HD) are undefined for empty targets—Dice vanishes, and HD returns infinite values. To reduce the number of false positives, we introduced a Thresholded Average loss, computed only over a set of matched controls, to minimise susceptibility artefact-induced errors. We averaged, over the controls, any voxel-wise probabilities exceeding 0.5, yielding a scalar penalty on false positives to be added to the aforementioned losses. The value of this loss is thus bounded between 0.5 and 1 in case of false positives and defined as 0 when a control is correctly identified as such. We assigned a weight to this penalty in the overall loss in response to training performance.

### 2.2.5 Validation metrics

During training, we evaluated performance after every epoch using two different but complementary measures. The lesion segmentation task aims to minimise the difference between our ground truth (label) and the lesion predicted by the model. To quantify this difference, we used the Dice metric on the binary prediction (after applying a sigmoid as an activation function to the logits and a threshold of 0.5), providing an index of the intersection between the label and the ground truth. As mentioned above, the Dice score is insensitive with small (in size but not necessarily in clinical consequence) lesions. Hence, we also used the Hausdorff Distance in the 95% confidence interval[40], which quantifies the distance between the contour of surfaces—here, the surface of lesioned areas and the surfaces of the predicted lesions.

## 2.3 Training

### 2.3.1 Cross-Validation

We trained every model (UNet, SWIN-UNETR and SWIN-UNETR+Ctr) with 5-fold cross-validation, dividing the training set into five balanced splits of equal size (712 lesioned images and 1380 controls) and training the model five times, each time with a different split used as the validation set. The hardware used to train the models and their relatively large size (in the case of the SWIN-UNETR models) constrained the tractable batch size. We implemented gradient accumulation to maintain comparable batch sizes between experiments and align with common machine learning practices. Gradient accumulation consists of computing and summing model gradients over multiple smaller mini-batches and then averaging to simulate a larger batch size. The effective batch size for each model was then batch size times the number of Graphic Processing Units (GPU) times gradient accumulation delay. SWIN-UNETR+Ctr had an effective batch size of 32 (batch size: 2, GPUs: 4, gradient accumulation: 4), the other SWIN-UNETR 16 (1, 2, 8) and the U-Net 32 (16, 2, 1).

### 2.3.2 Data balancing method

The marked heterogeneity of infarct anatomical characteristics makes the division between folds challenging. We chose to balance the splits by morphological phenotype and lesion volume to reduce the chance of having a split containing a limited variety of lesions. We generated 50000 randomly shuffled permutations of the five splits, aiming to achieve a balanced number of images belonging to each of the phenotypes described by Bonkhoff et al.[8], where membership was defined by the phenotype with the largest Dice score with the lesion. To create compact and focused phenotype masks, we thresholded the voxel-wise average maps from the original paper at 10%, ensuring they captured the core regions of the archetypes. For each permutation, we computed a Kruskall-Wallis[41] test to evaluate how different the distribution of lesion volume is between the five folds—we chose the best permutation as having the highest p-value and the lowest variability in average volume and standard deviation of volume across the splits.

### 2.3.3 Early stopping

To avoid overfitting and maximise performance, we implemented early stopping, activated once model performance plateaued. Empirically, we determined that the training could stop once the Dice and HD stopped improving together for more than 150 epochs. This threshold was chosen because, in all our tests, models that did not show improvement within 150 epochs failed to show any significant improvement even after 500 epochs or more. Combining Dice and HD as an early stopping criterion prevents the training from overfitting by minimising only the Dice criterion as the Focal Loss decreases. We selected the best model with the best Dice score between the last epoch in which both Dice and HD improved and 150 further epochs.

## 2.4 Evaluation methods

### 2.4.1 5-fold cross-validation

As explained above, we used 5-fold cross-validation for each model type (SWIN-UNETR+Ctr, SWIN-UNETR, U-Net). To maximise the value of our dataset and explore the robustness of the training methods, we evaluated the performance of each model type on the entire dataset. For a given model, we can evaluate the performance on the part of the dataset that was discarded from the training; by combining the performance of the five folds, we obtain an out-of-sample performance for the whole 3563 training dataset. This strategy avoids the selection biases of a held-out testing set without sacrificing learning potential, as no hyperparameter tuning or model selection was performed during the cross-validation process. Instead, our goal is to demonstrate the consistency of performance across folds, highlighting the effectiveness of the architecture and augmentations.

### 2.4.2 Anatomical model calibration

Ischaemic strokes vary in the frequency of their anatomical phenotypes. Sampling bias from the imbalance of anatomical phenotypes is therefore inevitable. It is therefore essential to quantify the variation in fidelity with anatomical features. A simple approach is to perform voxel-wise mass-univariate analysis, where each tested lesion is labelled by the achieved performance score (Dice or HD) on the test sets of the 5-fold cross-validation paradigm. We undertook this in SPM. Lesions were first smoothed with an 8mm full-width half maximum (FWHM) Gaussian to facilitate multiple comparison corrections within SPM's statistical framework. We then fitted a voxel-wise general linear model (GLM) for each of the three candidate models (SWIN-UNETR+Ctr, SWIN-UNETR, U-Net) with lesion density as the dependent variable and Dice or Hausdorff Distance as the independent variable. To these univariate GLMs, we added multivariate GLMs, where lesion volume was included as a nuisance covariate, controlling for lesion volume. The resultant T statistical map, thresholded at a family-wise error (FWE)-corrected p-value < 0.05, reveals regions where the test metric is significantly associated with location, demonstrating variations in performance with anatomy. Note this mass-univariate approach does not disentangle the impact on the score of anatomical locations vs the lesion morphological distribution[22,42], but this is not necessary here, unlike the case of lesion-deficit mapping[43].

### 2.4.3 Morphological model calibration

To quantify the impact of lesion morphological features on variability in performance, we need to represent the anatomy of a lesion in a compact, inspectable, latent representational space[44]. Here, we used Uniform Manifold Approximation and Projection[45] (UMAP) to derive a 2D latent embedding of lesion morphological characteristics. We trained a UMAP (default number of neighbours of 15, a minimum distance between points of 0.1, and Euclidean distance metric) with all the ground truth labels. The UMAP latent space represents the voxel-wise characteristics of lesions in a 2D space where similar lesions are close together and dissimilar lesions are far apart, with the distance between points proportional to their differences. UMAP allows us to identify variations in segmentation performance across the space of compactly represented lesion patterns. We scaled the coordinates of the latent space between 0 and 1 and applied the same scaling to the models' embeddings to compare the distance between points.

### 2.4.4 False positive prediction on DWI-negative images

Although the utility of the models is strictly confined to segmentation—not diagnosis—and assumes a lesion is present, it is helpful to quantify resistance to erroneously segmenting susceptibility artefacts that may closely resemble real lesions. The inclusion of DWI-negative images in our dataset allows us to evaluate this ability for our three models. Along with the five-way split of the lesion dataset, we assigned each split with a fifth of the DWI-negative images to use as a validation set. In this way, even the model trained with DWI-negatives has not seen a fifth of the data for each fold, allowing a fair performance comparison. Furthermore, besides presenting the descriptive statistics of the number of false positive voxels for each model, we performed bootstrapping on the distribution to allow us to run parametric statistics and help visualise the differences between models.

### 2.4.5 Noise model calibration

Real-world images are corrupted by noise. The performance of a segmentation model should thus be quantified by its variation with the degree of corruption. Here, we introduce a method for such noise calibration. We empirically selected parameters for the bias field transformation and Gibbs and Rician noise—commonly seen in DWI images—at the limit of what would make most images uninterpretable. Then, we tested each model's performance, gradually increasing the intensity of the noises from no noise to the selected parameters. For each noise setting, we explored between 10 and 20 increments of the noise intensity. We increased the number of increments when we saw that the performance of the models could still significantly decline.

### 2.4.6 Data availability

All the Python, Bash, and MATLAB codes used to create the preprocessing pipeline are available at https://github.com/chrisfoulon/mri_preprocessing/tree/v0.2.1. The Python code used to train, segment and analyse the models and their performance and the models' weights are available at https://github.com/chrisfoulon/lesseg_unet/tree/v1.0.1 with dependencies from https://github.com/chrisfoulon/BCBlib/tree/v0.4.1.

# Results

## 3.1 Standard performance evaluation

We trained every model five times using a different dataset split as a validation set—there was no model update on the validation set. We quantified the differences between three different models, a U-Net without data augmentation, SWIN-UNETR with data augmentation, and SWIN-UNETR with the same augmentations and the addition of the control dataset (SWIN-UNETR+Ctr). The Dice loss + Focal loss in every model was below 0.1 at the end of the training (Supp. Fig. 1). Unsurprisingly, both SWIN-UNETR models outperformed the baseline U-Net (Avg. Dice 5-folds: 0.8408, HD: 3.9200 voxels) on every metric. SWIN-UNETR+Ctr slightly underperformed (Avg. Dice 5-folds: 0.8915, HD: 2.6636 voxels) SWIN-UNETR (Avg. Dice 5-folds: 0.8952, HD: 2.6498 voxels) on these standard metrics (Supp Fig. 2). Figure 2A shows the overlap of the ground truth labels along with the overlap of the predictions (Figure 2B) from SWIN-UNETR+Ctr with the result of all the folds combined.

## 3.2 Anatomical performance evaluation

The fidelity of an optimal model ought to be invariant to the anatomical location of lesioned tissue. Since anatomical locations vary widely in their sampling, this is difficult to achieve and ought to be explicitly quantified for any candidate model. Here, we introduce the use of mass-univariate spatial inference to quantify the relationship between voxel-wise lesion density and segmentation performance. An ideal model should show no systematic relationship across the brain. Mass-univariate mapping of the U-Net model performance showed a significant association of voxel-wise lesion mass with Dice in anatomical areas within perforating MCA infarct territories (family-wise error (FWE)-corrected left MCA: p=0.01; right MCA p=0.026) (Figure 4). This association was no longer significant with the addition of lesion volume as a covariate. Neither SWIN-UNETR nor SWIN-UNETR+Ctr showed any significant anatomical modulation, and none of the models with HD. Though failure to reject the null hypothesis does not prove it, the well-established high sensitivity of mass-univariate inference at this data scale provides reasonable assurance of equitable performance across brain anatomy.

## 3.3 Morphological performance evaluation

Ischaemic lesions exhibit highly characteristic morphological structures inadequately captured by anatomical location alone. To relate voxel-level morphological features to performance, we must embed them into a surveyable latent space[46]. We used two-dimensional UMAP embeddings trained on the ground truth lesion labels to encode their locations and morphological profiles into a latent space (Figure 5). The predicted lesions from each segmentation model were then embedded into this same space, with the embeddings scaled (between 0 and 1) and translated (to start at zero) using the scaling and translation calculated from the ground truth embeddings. This alignment ensured all embeddings were directly comparable. We observed greater systematic variation, particularly in centrally projected, heterogeneous morphologies, for the U-Net model compared to the others. Despite this, the average and median distances from the ground truth embeddings were similar across all models: 0.16649 and 0.16641 for SWIN-UNETR+Ctr and SWIN-UNETR, and 0.17173 for the U-Net (median: 0.11120, 0.11693, 0.11159). However, when we statistically compare the distributions of distance—using paired t-tests—between models, we can observe a significant difference between the U-Net and SWIN-UNETR+Ctr (t-stat: 3.988, p-value < 0.0001) and between the U-Net and SWIN-UNETR (t-stat: -4.09, p-value < 0.0001). We found no significant differences between the two SWIN-UNETR models (t-stat: 0.069, p-value: 0.94).

## 3.4 False positive control rate evaluation

Since we trained all three models on the same data—each fold using the same split of the dataset for validation—we can use the split of the control data of each fold as a validation set to quantify the number of voxels in the control image erroneously predicted as lesioned, thus giving us a false-positive rate. In doing so, even the model trained using the control images has never trained on these images, enabling performance comparison between models. SWIN-UNETR+Ctr predicted 773 images out of 6900 with false positives with an average of 3.31 voxels (29.57 for only non-empty images). On average, SWIN-UNETR and the U-Net predicted 126.80 (346.79) voxels and 140.17 (276.33) voxels, respectively, in 2523 and 3500 control images. We conducted a bootstrap statistical analysis—using 10000 subsamples of 100 randomly picked images—to explore and illustrate the differences in inter-model false-positive volume in more detail (Fig 6). A paired t-test (Table 1) shows a highly significant difference between the SWIN-UNETR+Ctr and the U-Net (t-stat: -23.32, p-value << 0.0001). This difference is an order of magnitude greater than that between SWIN-UNETR+Ctr and SWIN-UNETR (t-stat: -284.20, p-value << 0.0001) and between SWIN-UNETR and the U-Net (t-stat: -341.11, p-value << 0.0001).

## 3.5 Robustness to image noise

Clinical images vary widely in their corruption by noise of various kinds. Unless the impact of such corruption is explicitly quantified, differences in model robustness to noise in real clinical settings cannot be adequately captured. Here, we therefore examined the performance of our models as a function of parametrically modulated bias field, Gibbs noise, and Rician noise corruption (Figure 7), quantifying differences by paired t-tests between models' average performance measures (Table 2). U-Net consistently performs worse than both SWIN-UNETR models on all noise configurations, exhibiting significantly lower Dice score and higher HD (FDR-corrected p-value < 0.05). We also observe that SWIN-UNETR+Ctr performs significantly better than SWIN-UNETR on Dice when all the noises are combined and on HD for bias field noise. The advantage of SWIN-UNETR+Ctr is most prominent with Rician noise (FDR-corrected p-value < 0.0001). We also examined the volume of the predicted lesion as an index of the behaviour of each model under increasing image corruption. As the combined noise and the bias field increase, the U-Net predicts significantly more voxels than the other models. However, we do not observe a difference with Gibbs noise, but it predicts fewer voxels as more Rician noise is added. Finally, the SWIN-UNETR significantly predicts fewer voxels as the Gibbs noise increases but predicts more with the Rician noise than SWIN-UNETR+Ctr.

# Discussion

Stroke is unique amongst neurological disorders in combining massive global health impact, variable responsiveness to rapid treatment, and plausibly high sensitivity to the underlying neuroanatomical patterns of damage. These features ought to make it a prime target for automated dense lesion segmentation since predicting clinical outcomes—both natural and treatment-conditional—may reasonably be expected to depend on the quality and resolution of lesion characterisation. Yet the development and application of automated methods here have lagged behind other neurological disorders, such as neuro-oncology, neuroinflammation and neurodegeneration, where the clinical value of segmentation maps is arguably less compelling.

As outlined in the introduction, progress is obstructed by the challenging nature of the imaging signals, the segmentation task, and the reigning data regime. This has motivated our creation of the largest dataset of expert-validated, anatomically registered acute stroke lesions segmented from DWI and the development and evaluation of an array of 3D lesion segmentation models based on state-of-the-art architectures conceived with these challenges expressly in mind. To a comparatively simple U-Net baseline, we compare two sets of SWIN-UNETR models. The first combines the SWIN-UNETR architecture with a diverse data augmentation scheme, demonstrating clear superiority over the baseline across a comprehensive panel of performance metrics. The second introduces a set of lesion-free control images with modification of the loss function, theoretically enabling better rejection of the high-signal susceptibility artefact that commonly corrupts DWI. To our knowledge, both SWIN-UNETR models set the current state-of-the-art for ischaemic stroke segmentation trained and validated on large datasets[47].

Crucially, the models reach the 0.85 Dice coefficient threshold held to justify preference over human expert segmentation[48] and do so on real-world clinical data evaluated at an unprecedented scale. Be that as it may, we wish to argue that simple, population average performance metrics, even enriched to include Hausdorff Distance as consensus guidelines recommend[20], provide insufficient characterisation of any segmentation or other featurisation algorithm proposed for clinical use or fundamental research intended to inform it. To fidelity must be added invariance over lesion and study characteristics that should not, at least *a priori*, influence performance. Two models of the same average fidelity may show different variations across subpopulations of lesions, impacting the equity of downstream tasks such as predictive or prescriptive inference. Summary metrics of variance alone would be inadequate, for the downstream impact may depend on the specifics of the subpopulation. For example, since the functional impact of damage varies dramatically across the brain, the level of achieved fidelity in a given anatomical region may matter more than in another. Equally, a model may be more or less sensitive to scan quality that naturally varies widely in real-world clinical practice, with no means of repeating a time-sensitive study.

Here, we therefore introduce a new framework for evaluating lesion segmentation models of anatomically organised domains such as the brain. Our framework seeks to expose and quantify variation with anatomy—of both the lesions and the underlying brain—and data quality within the range expected to be obtained in real-world practice.

We first evaluate variation in segmentation fidelity with the anatomy of the lesioned brain. This is enabled by operating with non-linearly registered lesion maps that allow faithful voxel-wise comparison across a population. An ideal model would exhibit performance invariant to the anatomical location of damage, i.e. the lesion density (over the population) of every voxel would be equally unrelated to the metric of fidelity. We formalise this statistically by mass-univariate voxel-wise inference, evaluating a regression model of lesion density as a function of the chosen metric, with covariates such as lesion volume optionally added. We show that whereas the baseline U-Net model shows marked variation across the brain, with substantially better performance in the MCA territory than elsewhere, both SWIN-UNETR models are only weakly influenced by anatomy, perhaps owing to their theoretically greater expressivity for complex lesion patterns[31]. This anatomical variation may be summarised by averaging the statistics across the brain into a single scalar or reported regionally for demarcating anatomical areas of greater or lesser quality. The frequentist statistics used here, identifying voxels where the null hypothesis of no relation to fidelity is rejected, may be replaced by Bayesian posterior probability maps[49], allowing distributional interpretation of the parameters. Note that this analysis does not seek to infer dependence on anatomy independently of lesion morphology—that would require a multivariate model—but rather the anatomical distribution of joint effects for comparing models and delimiting their anatomical application.

Lesions are not adequately described by their anatomical location, modelled independently at every voxel, for they exhibit complex patterns of co-occurrence across voxels[8]. We therefore secondly evaluate variation in segmentation fidelity with the morphology of the lesioned brain: the multivariate pattern of co-occurrent damage. Since the ambient space of lesions is high-dimensional, this requires a structure-preserving low-dimensional latent representation that renders morphological variations intuitively inspectable and robustly measurable. Here we employed UMAP to project lesions into a 2D space, thereafter labelled by performance metrics. As before, an ideal model would exhibit performance invariant to the morphology of the lesion, i.e. the metric of fidelity would be the same across the representational space. We find, once again, that U-Net exhibits greater variation than either SWIN-UNETR models. Although we do not formalise the comparison statistically here, topological inference could be used to identify regions in the latent space where the variation is significant, and to demarcate morphological subtypes above or below a criterion defined by utility[50].  Note that this morphological analysis is naturally not independent of anatomy, for lesion morphology and anatomy inevitably interact. Since the same anatomical region, however, may be affected by distinct lesion morphologies with different relations to model fidelity, both aspects require examination.

We thirdly evaluate variation in segmentation fidelity with the quality of the source image. The constraints of real-world clinical practice impose a greater range of imaging characteristics across acquisition parameters, resolution, coverage, noise, and artefact than is common in the research world. Resistance to image corruption ought therefore be an essential index of performance. Since there is typically no ground truth for any real-world image, here we implement an approach based on relating performance to the magnitude of added noise plausibly encountered in clinical settings: Rician noise, bias field and Gibbs noise[51,52], alone and in combination. This approach enables finer discrimination between models, revealing SWIN-UNETR+Ctr to be more resilient to image corruption (for Rician noise) than SWIN-UNETR, from which it is otherwise hard to distinguish. Though implemented here with additive noise, this approach is extensible to any manipulation of signal quality, and is naturally framed within the framework of psychometric function estimation[53], to be developed in further work.

The task of lesion segmentation presupposes the existence of a lesion to segment. Where the diagnostic framing permits confident distinction between lesion-positive and lesion-negative images—i.e. discrimination from other causes of abnormal DWI signal is not required—lesion detection and segmentation may be fruitfully combined. Our SWIN-UNETR+Ctr model is trained on both types of image, with a custom Thresholded Average loss, resulting in substantially fewer false positives where no lesion is present in the context of identical or superior segmentation performance. This approach is replicable in any domain where the diagnosis has been securely constrained to a binary choice but is, of course, no substitute for a full diagnostic model where no such constraint is possible.

## Limitations

A cornerstone of our approach—use of large-scale labelled data—mandates use of manually curated and modified, rather than densely segmented, ground truth maps. Dense manual segmentation at this scale is infeasible in practice, and smaller datasets do not permit evaluation of generalisability across anatomical and lesion-morphological characteristics, leaving the question of model equity unanswerable. Given the substantial disagreement across human experts[3,54] in this task, the advantage of dense manual labels is difficult to predict, and there is, of course, no mechanism to quantify it. Since the utility of lesion maps is ultimately in downstream prediction, prescription, and inference, the best validation is arguably necessarily deferred and constitutionally limited by the influence of non-imaging factors on modelled outcomes.

Our analysis of model performance does not include publicly available datasets such as ISLES[55]. This is because such datasets typically omit the b0 image from which non-lesioned anatomical characteristics are optimally obtained, and because our focus is on real-world data of greater diversity than is observed in the research setting from which such collections are drawn.

To facilitate learning of the underlying anatomy, our approach operates on non-linearly registered images and presupposes their successful transformation into standard template space. Since we employ relatively simple, compact registration models extensively employed in this domain[56–59], including on lesioned images, this preprocessing step is both robustly implementable and potentially absorbed into the segmentation model in future development.

# Conclusion

In this work, we trained three deep-learning algorithms on the largest DWI dataset of delineated acute ischaemic stroke lesions known to us. We carefully crafted a training scheme, balancing the dataset across five cross-validation folds and selecting data augmentations and transformations to prevent the models from overfitting and improve generalisability. We demonstrated that the two trained SWIN-UNETR architectures could reach state-of-the-art for the task using this training scheme according to the performance measure classically used in the machine learning community. We developed the Thresholded Average loss and showed that it allows us to use images without ischaemic damage without impairing the segmentation performance of SWIN-UNETR+Ctr with reasonable noise levels. We developed four novel methods to evaluate and compare the performance of segmentation models to reflect their potential translatability to clinical use better. These analyses established that SWIN-UNETR+Ctr is more resilient to common noise sources in diffusion imaging and is less likely to over-segment lesions when it cannot distinguish them from artifactual MRI intensity. Therefore, when the data is available, this work suggests that using control images is beneficial for training segmentation models. We hope this work will help researchers build better machine-learning models for medical imaging.

# Acknowledgements

Supported by the EPSRC (2252409), the NIHR UCLH Biomedical Research Centre (NIHR-INF-0840) and Wellcome (213038).

# Inclusion & Ethics Statement

This study was performed under ethical approval by the West London & GTAC research ethics committee for consentless use of fully anonymized data. The data is an unselected sample based on clinical diagnosis of acute ischaemic stroke; no other specific inclusion or exclusion criteria were used. It should therefore be proportionally representative of the acute ischaemic stroke patient population presenting to hospitals in London, UK.

**Fig. 1: Dataset construction and curation process.** This flowchart summarises the curation and label correction steps from the raw datasets to the final lesion and control sets. Lesion images from Xu et al. (n=1333) and UCLH (n=5139) were reviewed by experts and semi-automated corrections, producing the final lesion dataset (n=3563). Control images from UCLH (n=6691) were manually curated (n=5900), and UK Biobank (n=1000) was randomly selected, forming the final control dataset (n=6900). An external lesion dataset (KCH, n=2674) was also included for lesion distribution validation.

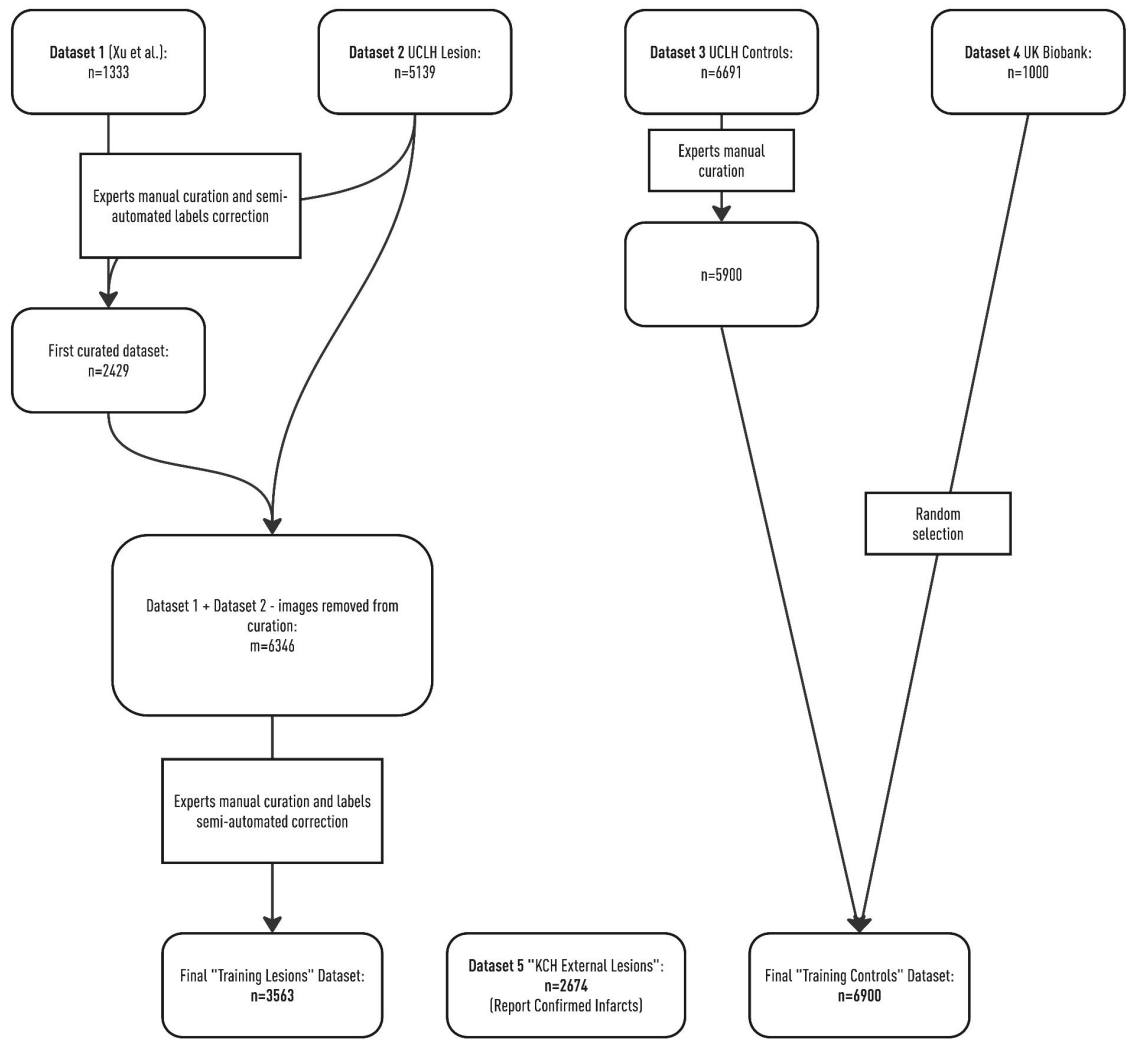

**Figure 2: Lesions overlap of the ground truths and SWIN-UNETR+Ctr predictions.** A) Overlap of the ground truth labels used for the training of the models. B) Overlap of the predictions of SWIN-UNETR+Ctr. The predictions of the different models are made on the same set of DWI images as the labels. The 5-fold cross-validation training scheme allows the prediction to be done on data unseen by the models.

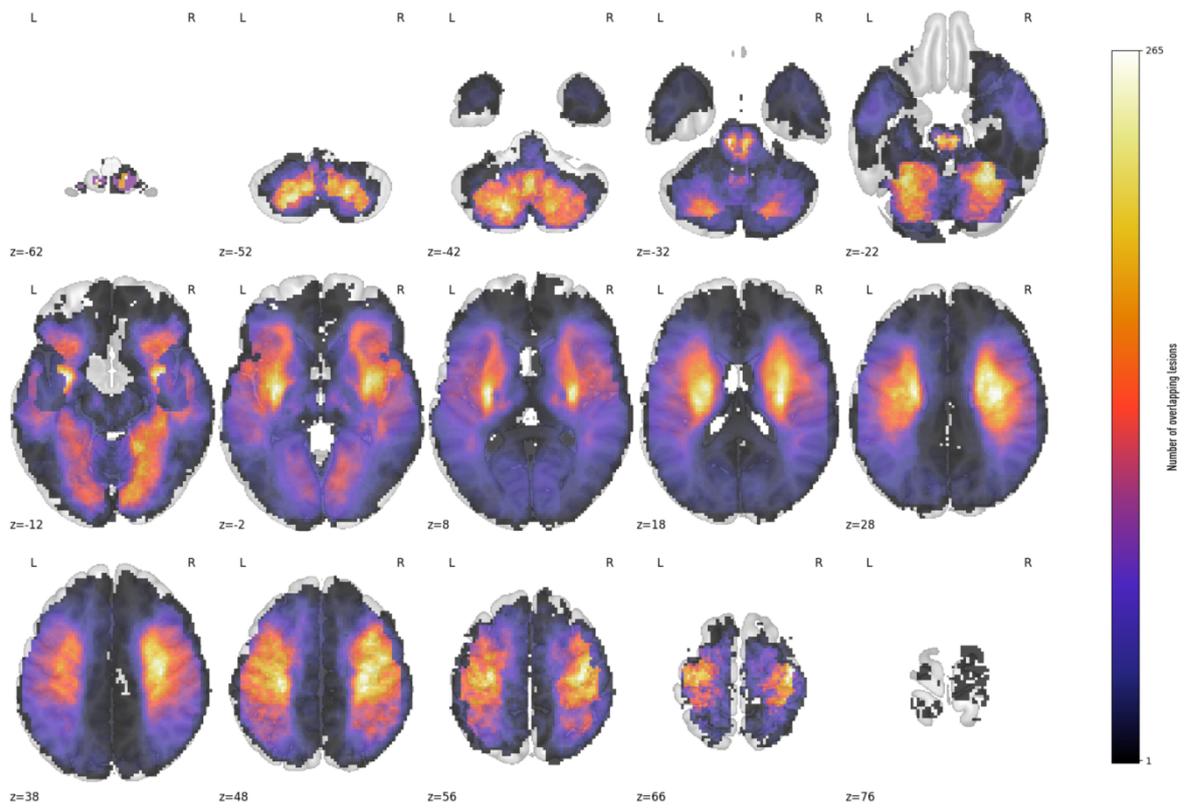

A) Ground Truth Labels

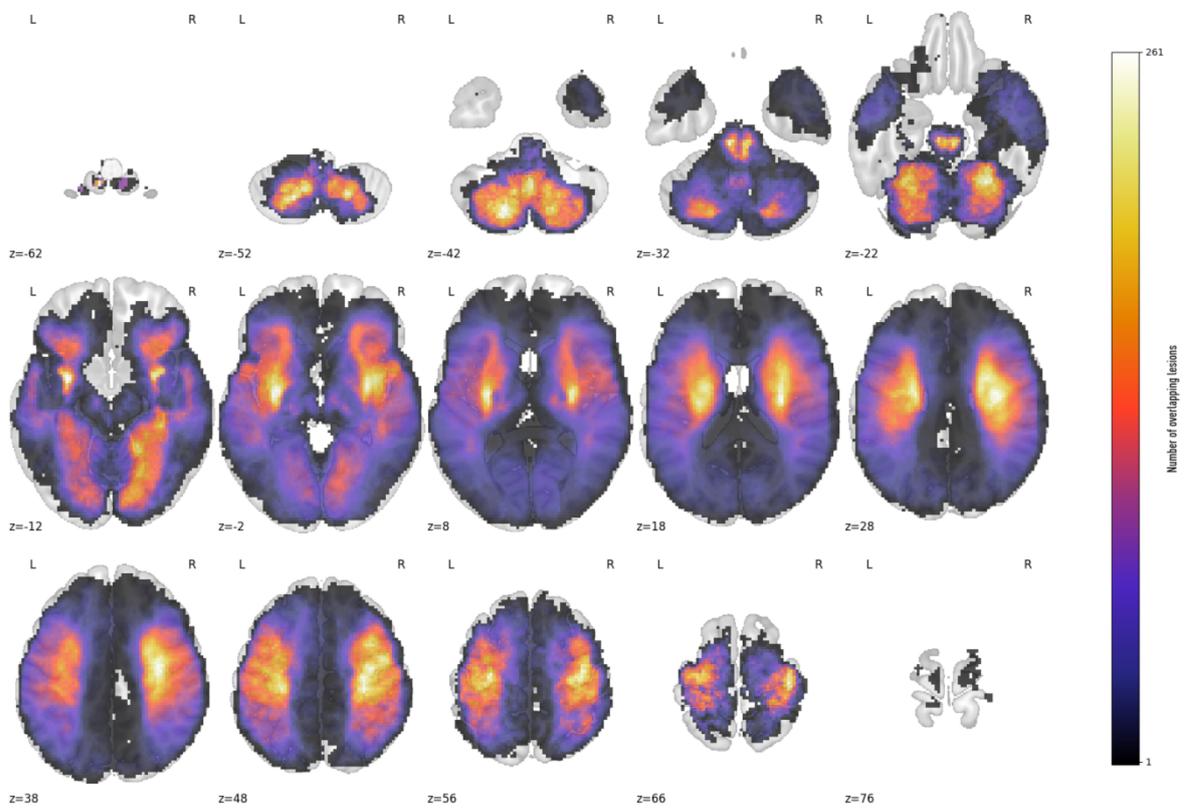

B) SWIN-UNETR+Ctr predictions

**Figure 3: Distribution of the predicted lesions of SWIN-UNETR+Ctr on the external dataset from the King's College Hospital.**

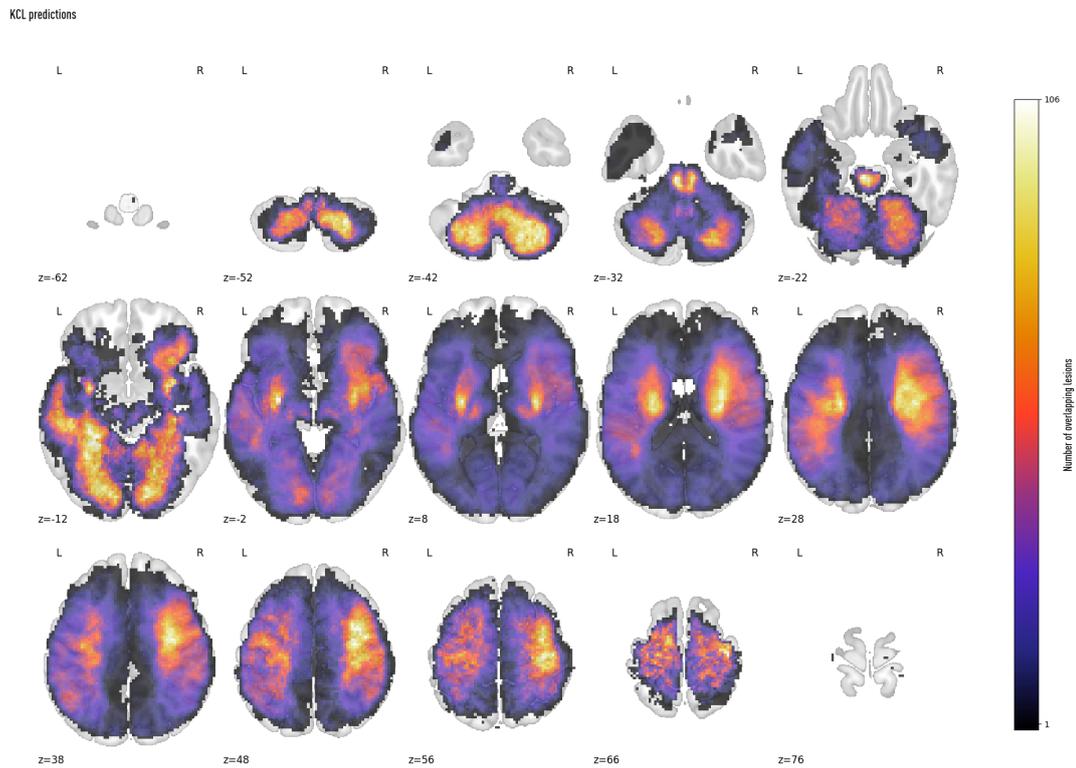

**Fig 4: Spatial performance inequity with U-Net models.** Clusters within bilateral middle cerebral artery territories denote clusters wherein significantly greater lesion segmentation performance (by Dice coefficient) is achieved with basic U-Net models (all FWE-corrected $p<0.05$). Note there was no significant performance inequity for either of the SWIN-UNETR models.

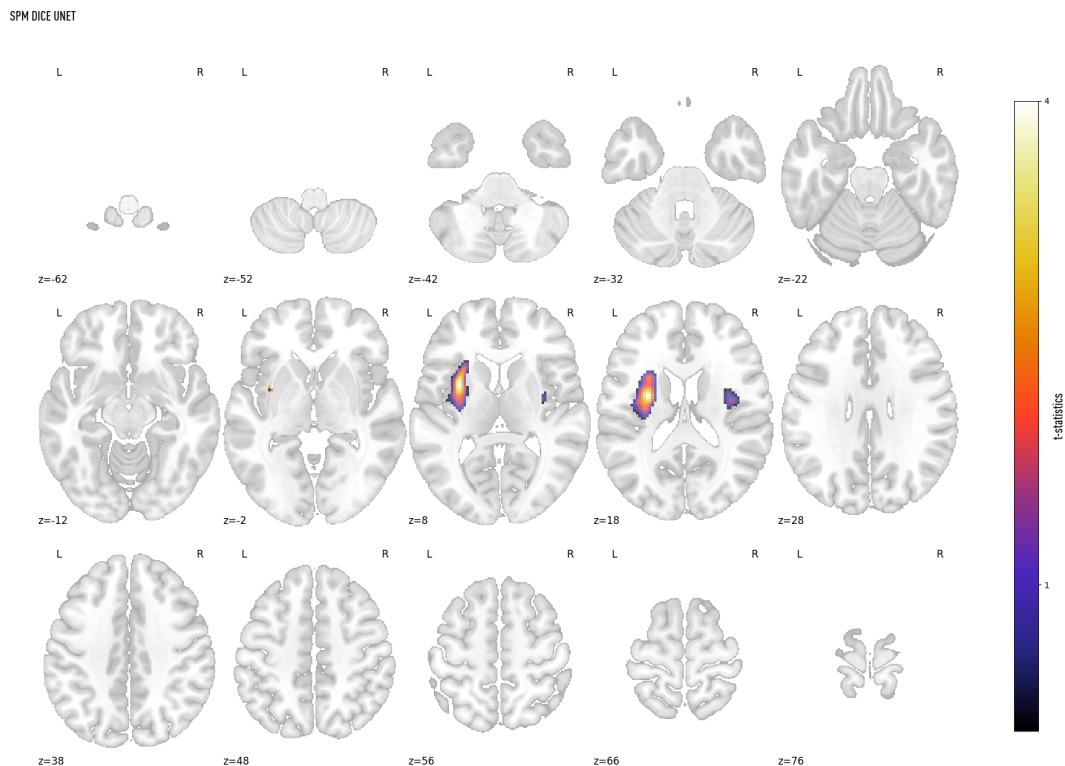

**Fig. 5: UMAP embeddings of lesion predictions for different models, evaluated with Dice metric and Hausdorff distance.** The top set of plots (A) represents the Dice metric between model predictions and ground truth labels, where lighter colours indicate better overlap. The bottom set of plots (B) represents the Hausdorff distance, with lighter colours indicating a smaller distance and better alignment of lesion boundaries. The bottom-left panel in each section shows the UMAP embedding of the ground truth labels, with marker size corresponding to lesion size in voxels. The other panels display the embeddings of lesions predicted by the three models: SWIN-UNETR+Ctr, SWIN-UNETR, and U-Net. This visualisation highlights the structural similarities and differences between predicted and actual lesion distributions across models, with variations in segmentation quality captured through Dice and Hausdorff metrics.

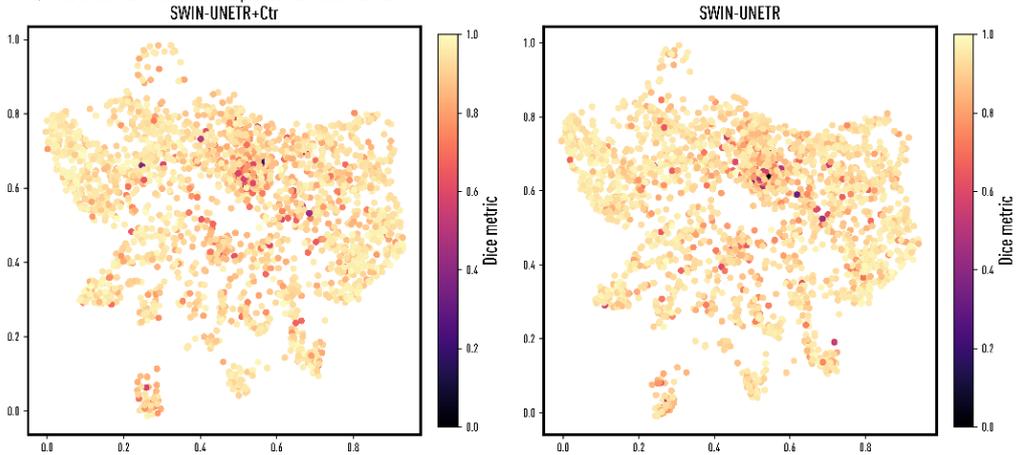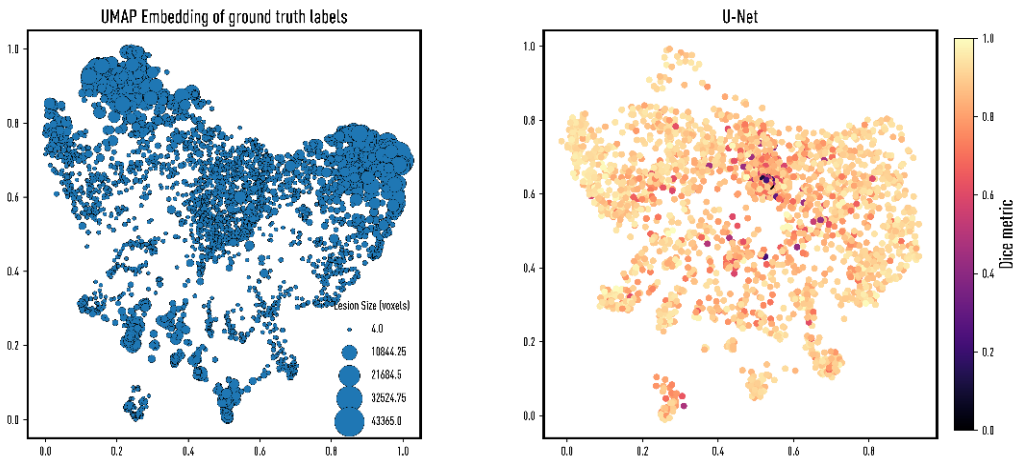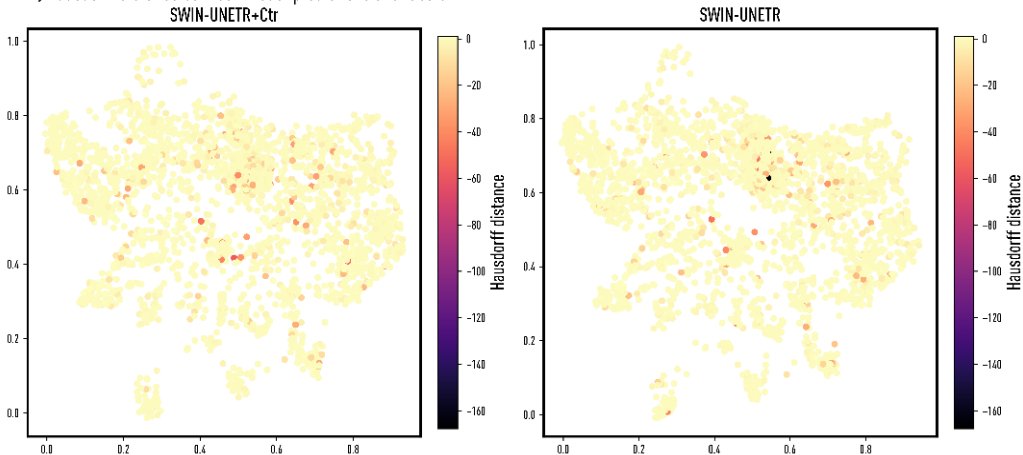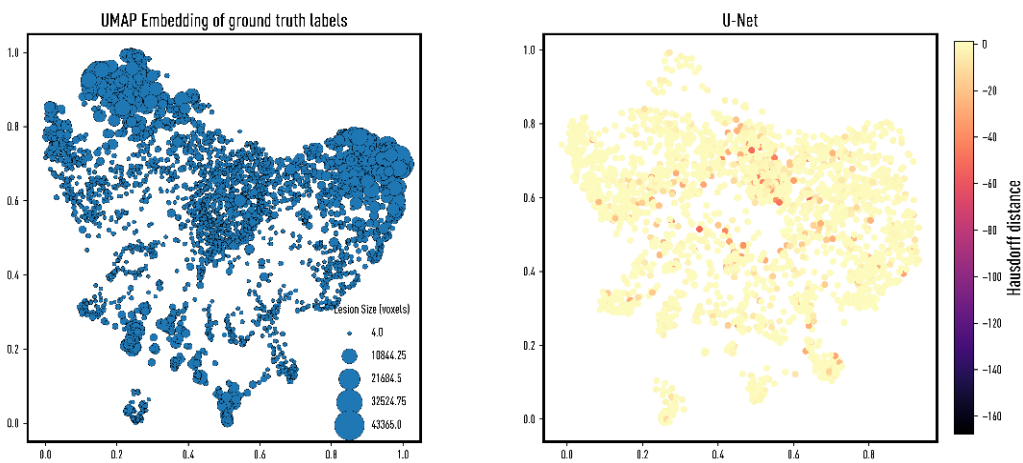

**Fig 6: Bootstrap distributions of out-of-sample means (number of voxels) of false positive volume on the control images.**

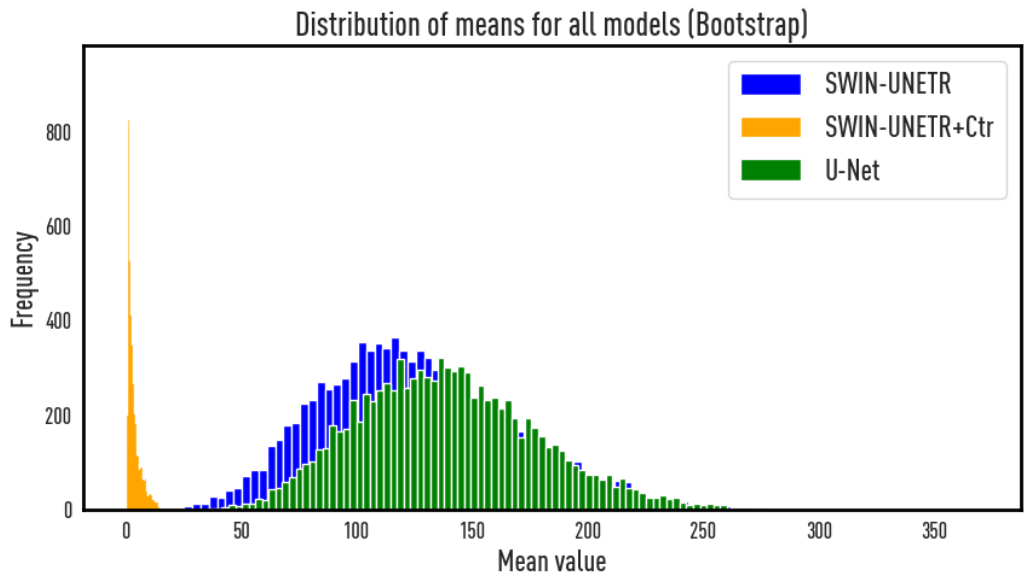

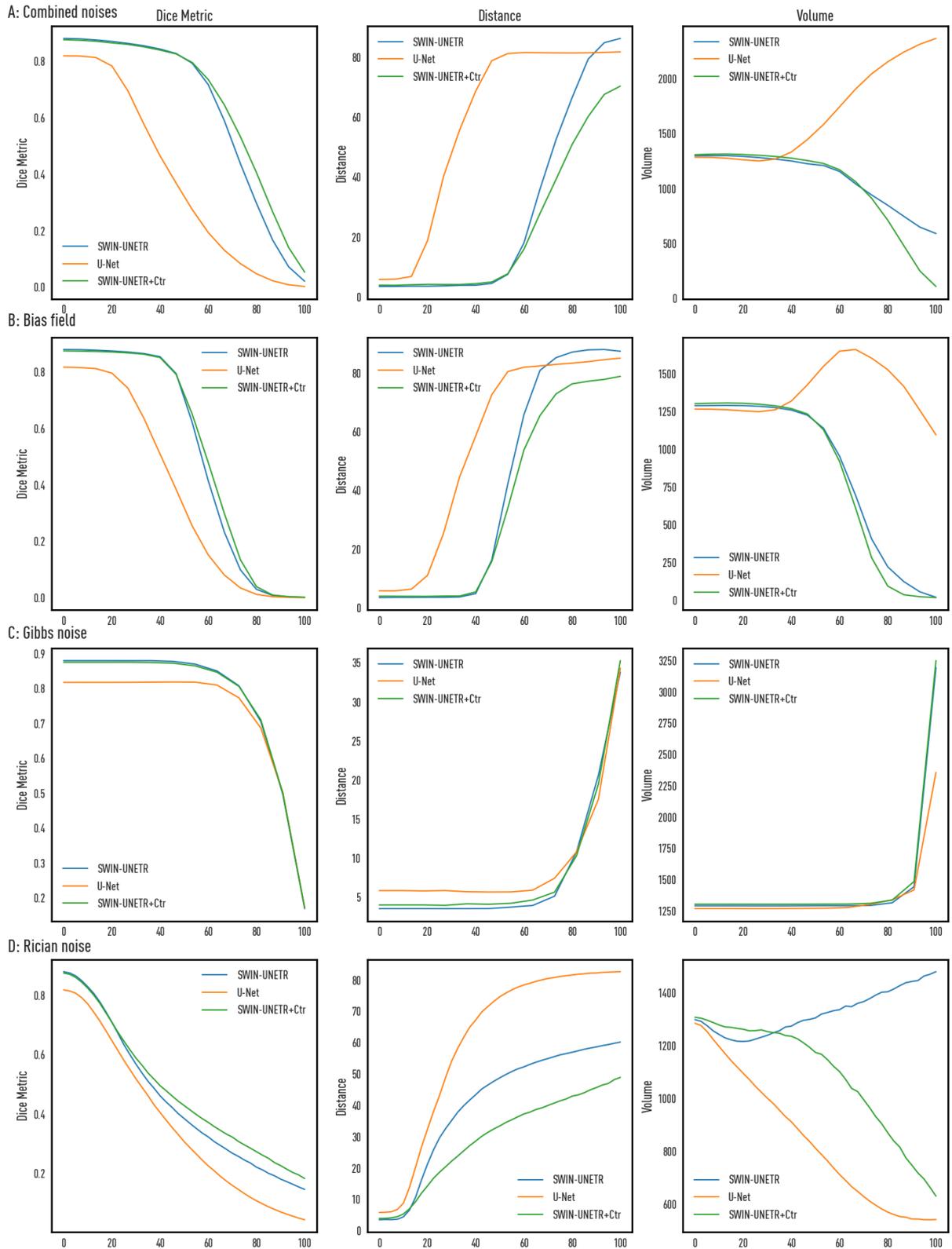

**Fig 7: Evolution of models' performance with increasing noise.** A) Combination of bias field, Gibbs noise and Rician noise from no noise to a noise level that make most images uninterpretable by an expert. B) Bias field increments. C) Gibbs noise increments. D) Rician noise increments

**Table 1: Descriptive statistics of the number of false positive predicted voxels per model on the control dataset. Bold values are the best values per measure.**

| Number of false positive voxels | SWIN-UNETR+Ctr | SWIN-UNETR | U-Net |
|---|---|---|---|
| Mean | **3.31** | 126.80 | 140.17 |
| Mean (non-zero) | **29.57** | 346.79 | 276.33 |
| Standard deviation | **34.22** | 434.83 | 401.35 |
| Standard deviation (non-zero) | **98.38** | 663.94 | 529.08 |
| Number of images with $\geq 1$ | **773** | 2523 | 3500 |
| Maximum in one image | **1114** | 6482 | 4920 |

**Table 2: Comparison of models across different noise types and metrics.** Bold values are significant with FDR correction. Values followed with * are significant after FWER correction.

Table 2: Comparison of Models across Different Noise Types and Metrics

| Noise | Metric | SWIN-UNETR > U-Net | | | SWIN-UNETR > SWIN-UNETR+Ctr | | | U-Net > SWIN-UNETR+Ctr | | |
|---|---|---|---|---|---|---|---|---|---|---|
| | | t-stat | FDR | FWER* | t-stat | FDR | FWER* | t-stat | FDR | FWER* |
| allnoises | Dice | 5.3180* | **0.0002** | 0.0021* | -2.6194 | **0.0240** | 0.1467 | -5.7489* | **0.0001** | 0.0010* |
| allnoises | HD | -4.0490* | **0.0020** | 0.0203* | 2.7781 | **0.0190** | 0.1443 | 5.3361* | **0.0002** | 0.0021* |
| allnoises | Volume | -3.4486* | **0.0056** | 0.0490* | 1.6766 | 0.1210 | 0.3309 | 3.1733 | **0.0094** | 0.0789 |
| bias | Dice | 4.0615* | **0.0020** | 0.0203* | -1.8278 | 0.0955 | 0.3309 | -4.1676* | **0.0019** | 0.0172* |
| bias | HD | -2.6701 | **0.0225** | 0.1467 | 3.5405* | **0.0049** | 0.0463* | 4.0564* | **0.0020** | 0.0203* |
| bias | Volume | -3.6449* | **0.0043** | 0.0399* | 1.9290 | 0.0820 | 0.3309 | 3.5421* | **0.0049** | 0.0463* |
| gibbs | Dice | 5.7284* | **0.0003** | 0.0030* | 2.3048 | **0.0500** | 0.2577 | -6.2215* | **0.0002** | 0.0017* |
| gibbs | HD | -3.0668 | **0.0154** | 0.1213 | -2.0517 | 0.0752 | 0.3309 | 2.9082 | **0.0190** | 0.1443 |
| gibbs | Volume | 1.2101 | 0.2516 | 0.3309 | -5.0192* | **0.0009** | 0.0086* | -1.4363 | 0.1838 | 0.3309 |
| rician | Dice | 21.1932* | **0.0000** | 0.0000* | -8.9285* | **0.0000** | 0.0000* | -16.4873* | **0.0000** | 0.0000* |
| rician | HD | -15.4076* | **0.0000** | 0.0000* | 12.3993* | **0.0000** | 0.0000* | 14.2794* | **0.0000** | 0.0000* |
| rician | Volume | 9.5807* | **0.0000** | 0.0000* | 5.2396* | **0.0000** | 0.0002* | -13.5663* | **0.0000** | 0.0000* |

# Supplementary material

**Supplementary Table 1: Descriptive statistics for each fold, including mean and standard deviation (SD) of patient age, proportion of females, and label size.** SD indicates standard deviation within each fold.

| fold   | Mean Patient Age | SD Patient Age | Proportion Female (F) | SD Proportion Female | Mean Label Size | SD Label Size |
|--------|------------------|----------------|-----------------------|----------------------|-----------------|---------------|
| fold_0 | 66.18            | 15.04          | 0.41                  | 0.49                 | 1324.84         | 3149.72       |
| fold_1 | 67.42            | 14.92          | 0.39                  | 0.49                 | 1332.50         | 3073.91       |
| fold_2 | 67.13            | 15.34          | 0.43                  | 0.49                 | 1323.92         | 3109.75       |
| fold_3 | 67.27            | 15.27          | 0.45                  | 0.50                 | 1336.89         | 3103.91       |
| fold_4 | 66.70            | 15.59          | 0.46                  | 0.50                 | 1319.78         | 2992.60       |

**Supplementary Figure 1: Training losses of the different models.** The training loss combines Dice Loss and Focal Loss, balancing overlap overlap accuracy and voxel-wise misclassification.

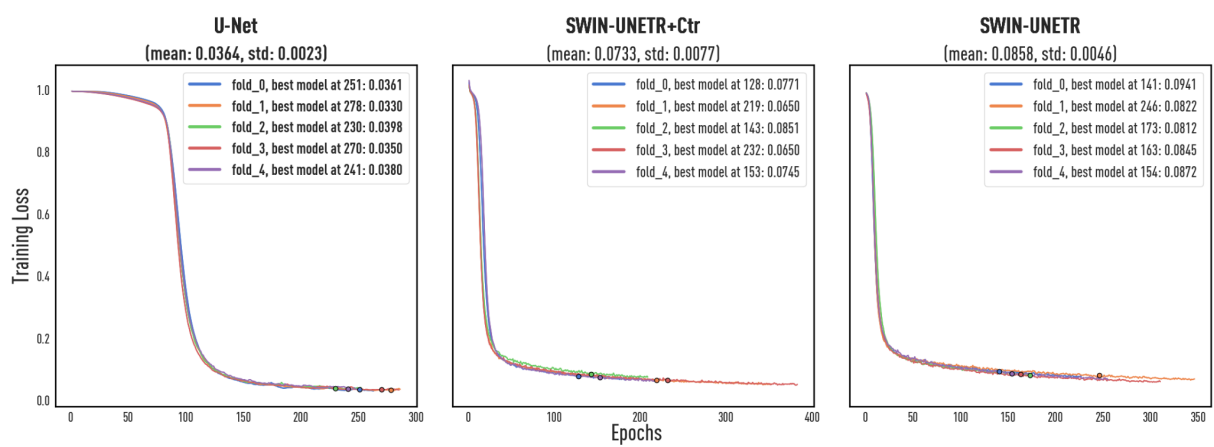

**Supplementary Figure 2: Descriptive statistics of each training epoch and of selected best models.** Top row shows the Dice score, on the validation set of each fold. Bottom row show the Hausdorff Distance on the validation set of each fold. The X-axes are the epoch numbers. On top of each plot is the mean and standard deviation of the average value of each fold. The best values of each row are displayed in bold.

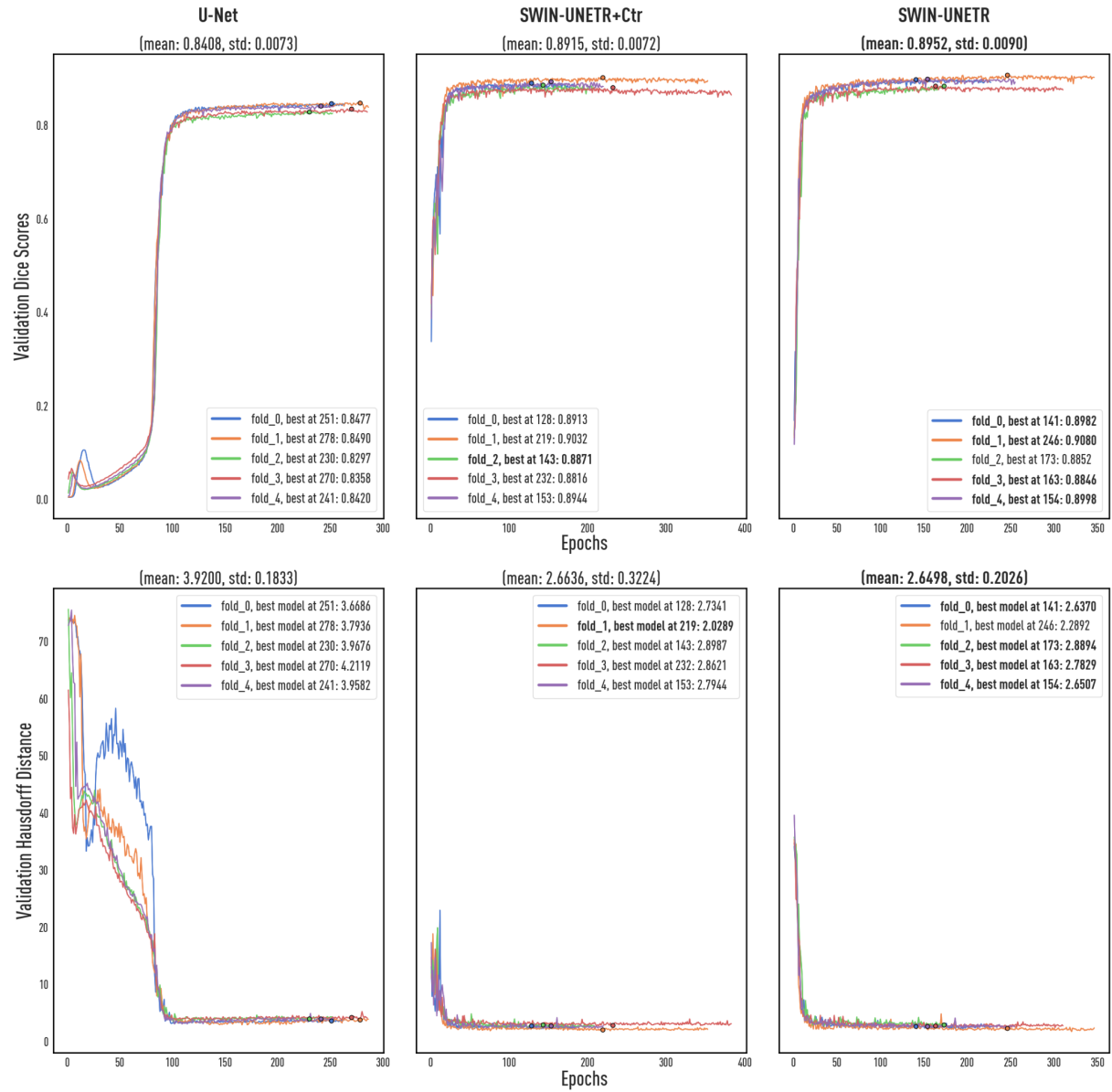